\documentclass[letter]{emulateapj}
\usepackage{natbib}
\usepackage{amsmath}
\usepackage{arydshln}
\usepackage{multirow}
\usepackage{rotating}

\begin{document}

\title{Discovery of Five New Pulsars in Archival Data}

\author{M.~B.~Mickaliger\altaffilmark{a}, D.~R.~Lorimer\altaffilmark{a,b}, J.~Boyles\altaffilmark{a,c}, 
M.~A.~McLaughlin\altaffilmark{a,b}, A.~Collins\altaffilmark{a}, L.~Hough\altaffilmark{a}, N.~Tehrani\altaffilmark{a}, 
C.~Tenney\altaffilmark{a}, A.~Liska\altaffilmark{a}, \& J.~Swiggum\altaffilmark{a}}

\altaffiltext{a}{Department of Physics, West Virginia University, Morgantown, WV 26506}
\altaffiltext{b}{Also adjunct at the National Radio Astronomy Observatory, Green Bank, WV 24944}
\altaffiltext{c}{Present address: Department of Physics, Western Kentucky University, Bowling Green, KY 42101}

\begin{abstract} 

Reprocessing of the Parkes Multibeam Pulsar Survey has resulted in the discovery of five previously unknown pulsars and 
several as-yet-unconfirmed candidates. PSR J0922$-$52 has a period of 9.68~ms and a DM of 122.4~pc cm$^{-3}$. PSR 
J1147$-$66 has a period of 3.72~ms and a DM of 133.8~pc cm$^{-3}$. PSR J1227$-$6208 has a period of 34.53~ms, a DM of 
362.6~pc cm$^{-3}$, is in a 6.7 day binary orbit, and was independently detected in an ongoing high-resolution Parkes 
survey by Thornton et al.\ and also in independent processing by Einstein@Home volunteers. PSR J1546$-$59 has a period of 
7.80~ms and a DM of 168.3~pc cm$^{-3}$. PSR J1725$-$3853 is an isolated 4.79-ms pulsar with a DM of 158.2~pc cm$^{-3}$. 
These pulsars were likely missed in earlier processing efforts due to the fact that they have both high DMs and short 
periods, and also the large number of candidates that needed to be looked through. These discoveries suggest that further 
pulsars are awaiting discovery in the multibeam survey data. 

\end{abstract}


\section{Introduction}

While targeted searches have been useful in finding unique pulsars, most pulsars known today have been found in 
large-scale, blind pulsar surveys. One such survey, the Parkes Multibeam Pulsar Survey \citep[PMPS;][]{Manchester:2001}, 
surveyed a strip along the Galactic plane using the 13-beam receiver on the Parkes 64-m telescope. Initial processing of 
the data resulted in the discovery of 742 pulsars \citep{Manchester:2001, Morris:2002, Kramer:2003, Hobbs:2004, 
Faulkner:2004, Lorimer:2006}. Another 44 pulsars and 30 RRATs (rotating radio transients) were found in further 
reprocessings \citep{Eatough:2009, Eatough:2010, Eatough:2011, Keith:2009, McLaughlin:2006, Keane:2010, Keane:2011}, and an 
additional 21 pulsars have been found so far by 
Einstein@Home\footnote{\url{http://einstein.phys.uwm.edu/radiopulsar/html/PMPS\_discoveries/}} (Knispel et al., in prep). 
The 44 additional pulsars were found due to the implementation of new techniques for removing terrestrial interference and 
new techniques for sorting pulsar candidates.

In this paper, we present the discovery of a further five pulsars in the PMPS data. The motivation for our re-analysis of 
the PMPS data was a single-pulse study, which will be presented elsewhere. Single-pulse studies involve the search for and 
characterization of transient, non-periodic bursts. We performed periodicity searches as well as single-pulse searches 
since the additional processing time was negligible. In Section \ref{data}, we describe the data reduction and analysis. 
Section \ref{psrs} details the five pulsars that we discovered, and conclusions are given in Section \ref{conc}.

\section{Data Reduction}
\label{data}

\begin{figure}[ht]
\includegraphics[scale=0.65]{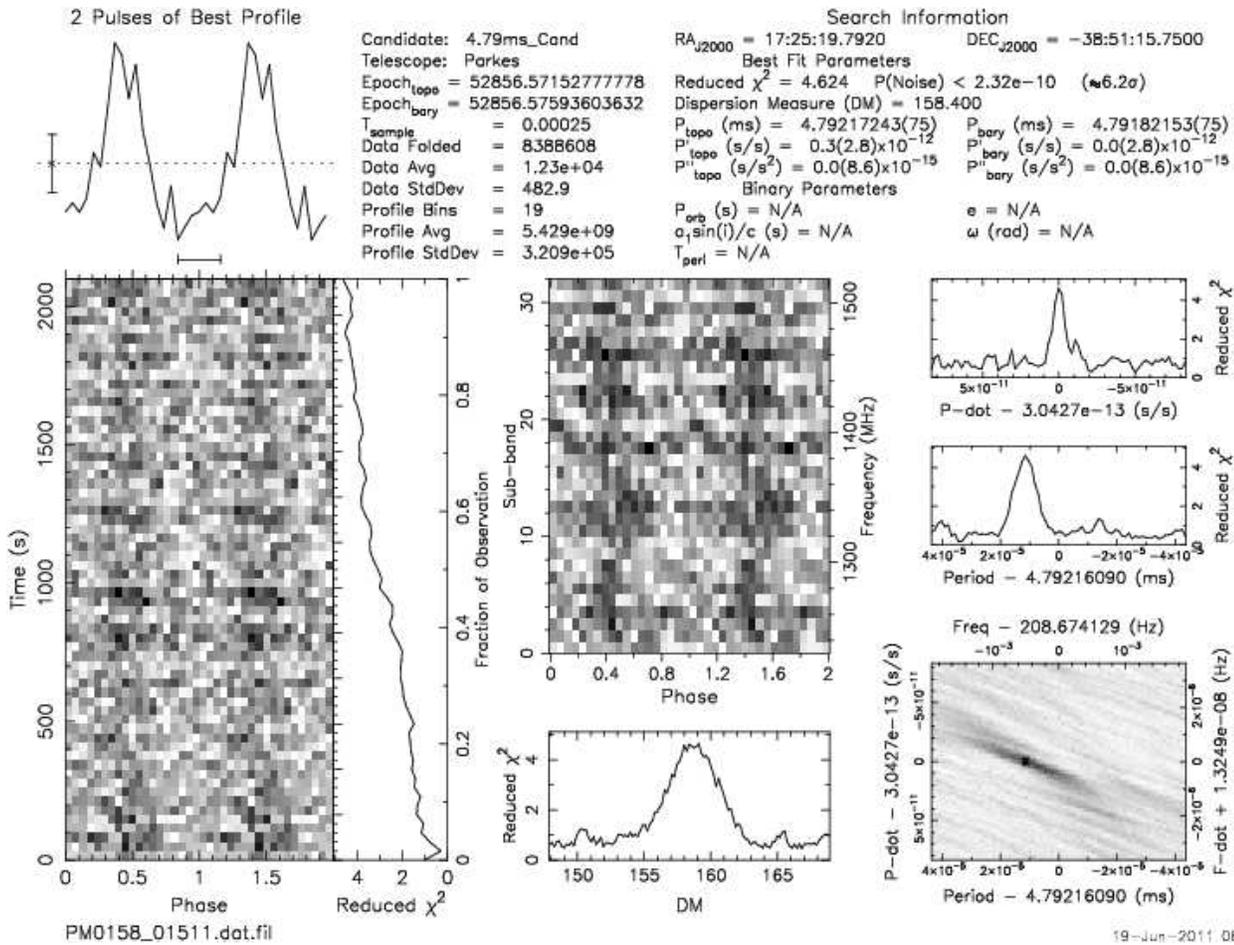}
\caption{A diagnostic plot from {\tt prepfold} showing the discovery of PSR J1725$-$3853. For a description of the 
subplots, see Section \ref{data} of the text.}
\label{fig:prepfold}
\end{figure}

The PMPS data were searched for periodic signals using freely-available analysis 
software\footnote{\url{http://sigproc.sourceforge.net}}. First, the frequency channels in the data were shifted to correct 
for the dispersion due to free electrons in the interstellar medium. This dedispersion is done at many dispersion measures 
(DM, which is the integrated column density of electrons along the line of sight) and results in a time series for each DM. 
The total number of DMs searched was 203 and was optimally chosen by the 
{\tt dedisperse\_all}\footnote{\url{http://www.github.com/swinlegion}} program, which we used for dedispersion due to its 
speed and efficiency. The time series were processed by {\tt seek}, a program that searches for periodic signals from an 
object. We searched for both periodic signals and single pulses out to a DM of 5000~pc cm$^{-3}$. This upper limit was 
chosen in order to be sensitive to highly-dispersed bursts. Results from the single-pulse search will be 
presented in another paper. The periodicity-search analysis method implemented in {\tt seek} is the standard Fourier-based 
approach \citep[see, e.g.,][]{Lorimer:2005} where the amplitude spectrum is subject to multiple harmonic folds, summing 2, 
4, 8, and 16 harmonics. This process increases sensitivity to narrow pulses in a close-to-optimal fashion 
\citep{Ransom:2002}. All candidate signals, with spectral S/Ns greater than six, are sought and saved during this process. 
After all DM trials have been searched, statistically significant candidates with S/N greater than nine are subject to 
further analysis. For these candidates, the raw multi-channel data are then folded at the period of each candidate using 
{\tt prepfold}, part of the PRESTO package\footnote{http://www.cv.nrao.edu/$\sim$sransom/presto}. The {\tt prepfold} 
program carries out a search to optimize the period from {\tt seek} and produces a set of diagnostic plots for each 
candidate. Figure \ref{fig:prepfold}, which is the discovery plot for PSR J1725$-$3853, is one example of these diagnostic 
plots. These plots consist of the following parts: an integrated pulse profile (upper left), which is the result of folding 
the data at the period of the candidate; a plot of pulse phase vs observation time (left), which shows the phase of the 
pulses as they arrive throughout the observation, as well as accumulated $\chi^2$ vs observation time \citep[for the 
definition of $\chi^2$, see Equation 7.3 in][]{Lorimer:2005}; pulse phase vs frequency (middle), which shows the phase of 
the pulses across the observing band; period vs $\chi^2$ (middle right), which shows the $\chi^2$ value resulting from 
folding the data at many trial periods; period derivative vs $\chi^2$ (upper right), which shows the $\chi^2$ value 
resulting from folding the data with many trial period derivatives; DM vs $\chi^2$ (lower middle), which shows the $\chi^2$ 
resulting from dedispersing the data at many trial DMs; and period vs period derivative (lower right), which shows the 
$\chi^2$ intensity in period$-$period derivative space.
 
No acceleration searches were carried out in this reduction. To reduce the number of plots that needed to be 
inspected by eye, we selected candidates with periods under 50~ms, DMs greater than 10~pc cm$^{-3}$, and spectral S/Ns 
greater than nine for viewing, resulting in $\sim$270000 candidates, five of which have already been confirmed as pulsars. 
Since this processing was never intended to be rigorous, we assumed that most pulsars with periods greater than 50~ms had 
been discovered, and that most candidates with a DM that peaked under 10~pc cm$^{-3}$ were interference.

\section{Newly Discovered Pulsars}
\label{psrs}

From our inspection of the {\tt prepfold} plots, we identified five very promising pulsar candidates and have subsequently 
been able to confirm these as new pulsars and perform follow-up observations as described below.

\subsection{PSR J0922$-$52}
\label{0922}

\begin{figure}[ht]
\centering
\begin{tabular}{cc}
\includegraphics[scale=0.25, angle=-90]{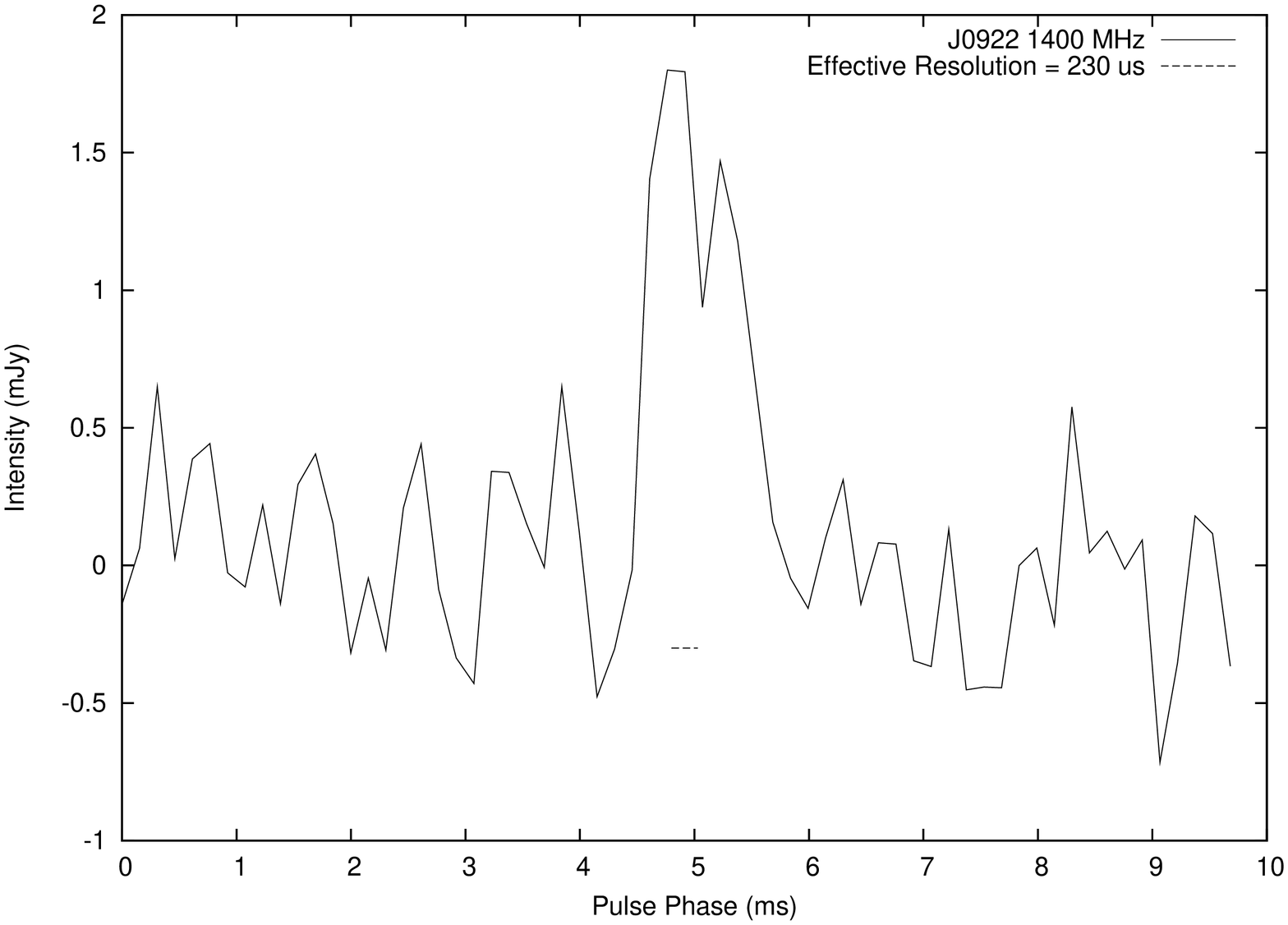} &
\includegraphics[scale=0.25, angle=-90]{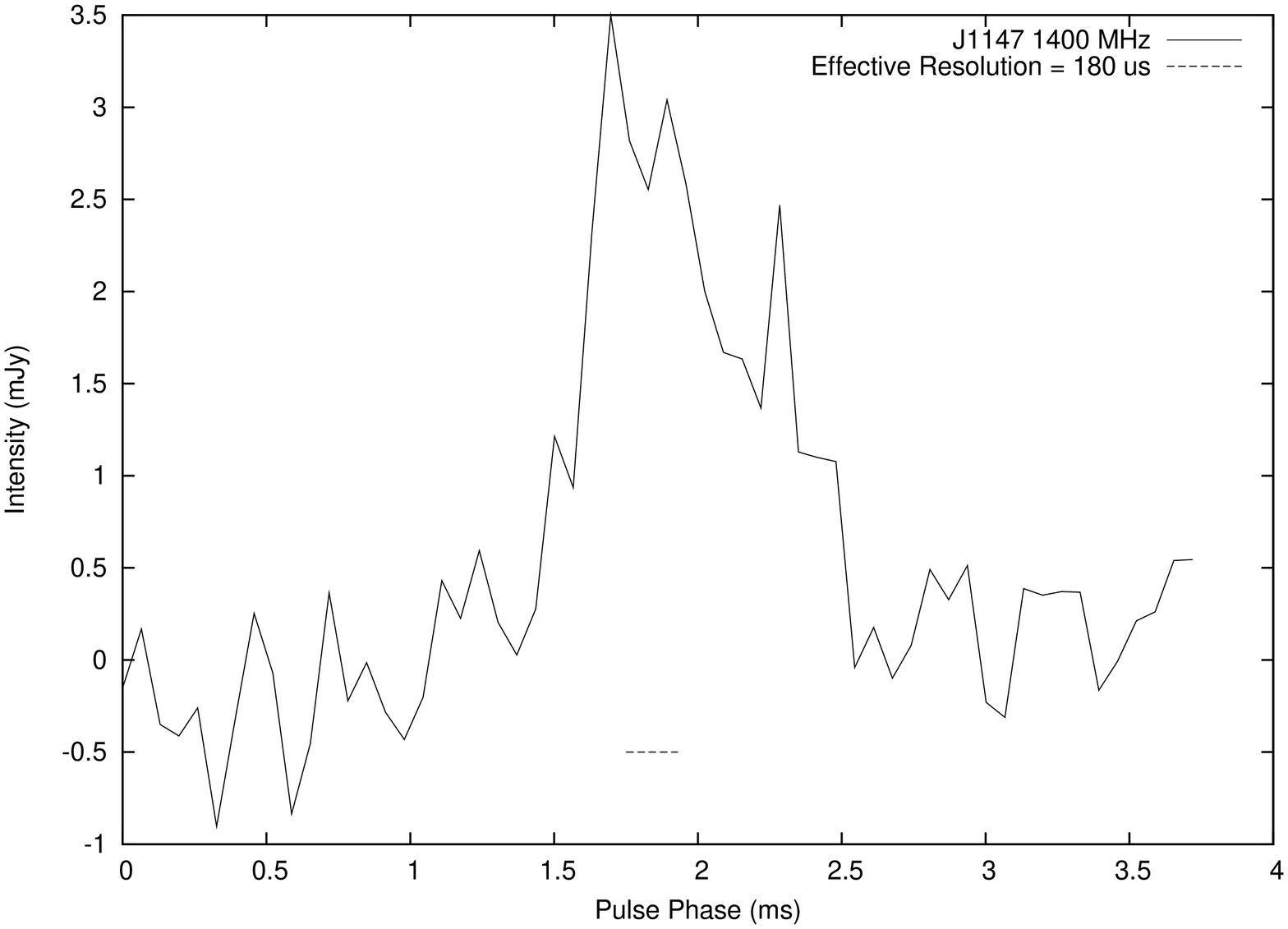} \\
\includegraphics[scale=0.25, angle=-90]{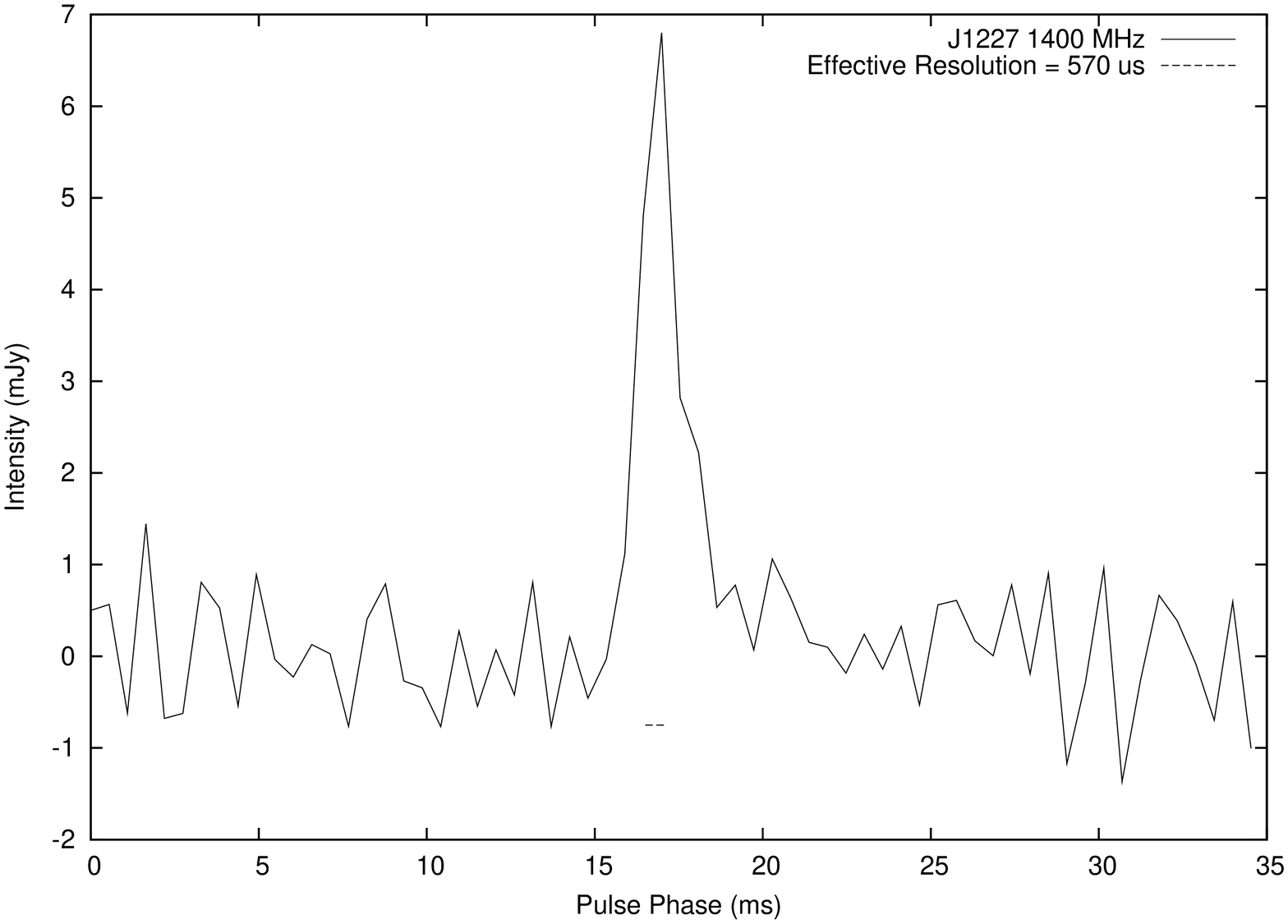} &
\includegraphics[scale=0.25, angle=-90]{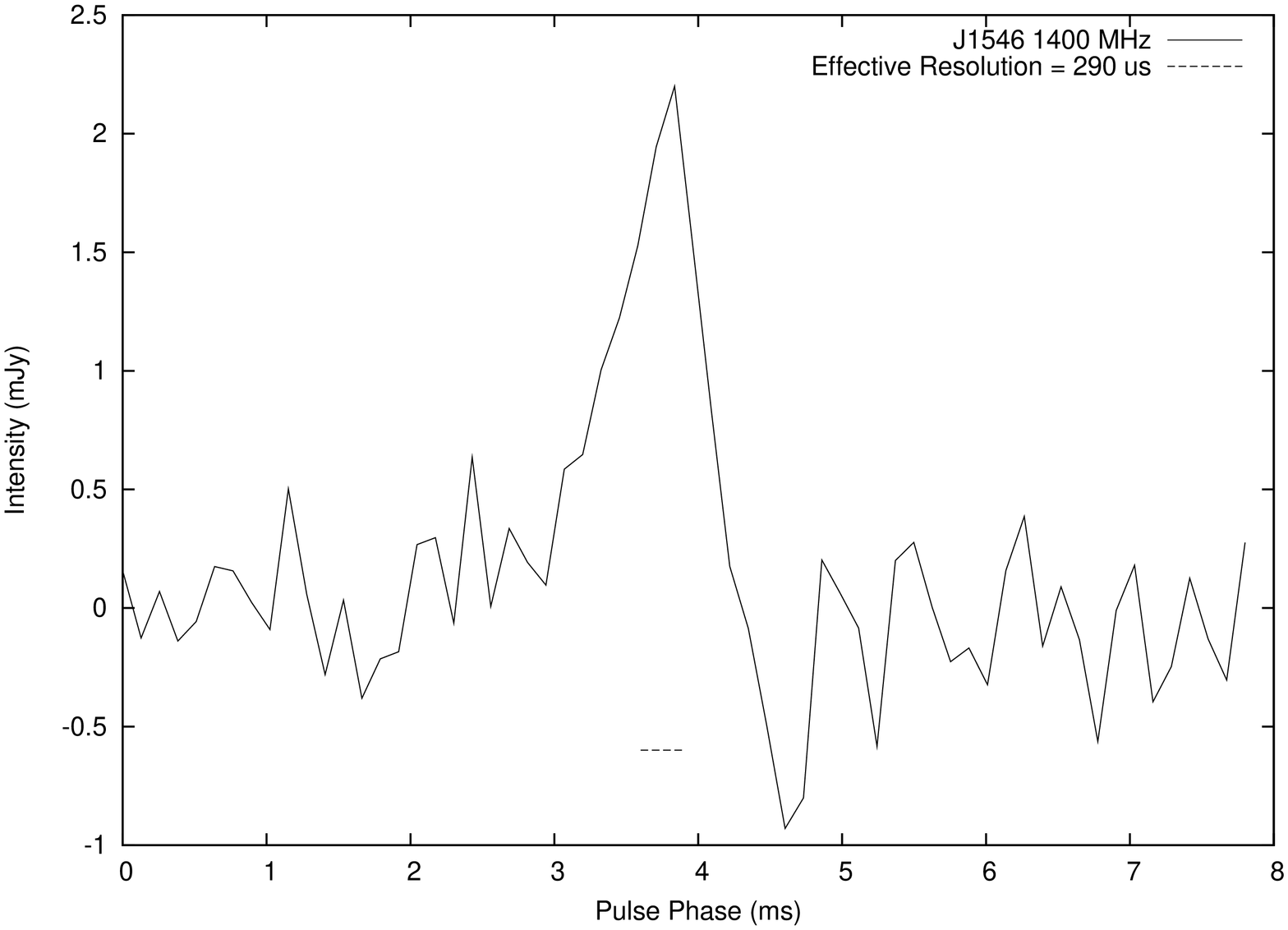} \\ 
\includegraphics[scale=0.25, angle=-90]{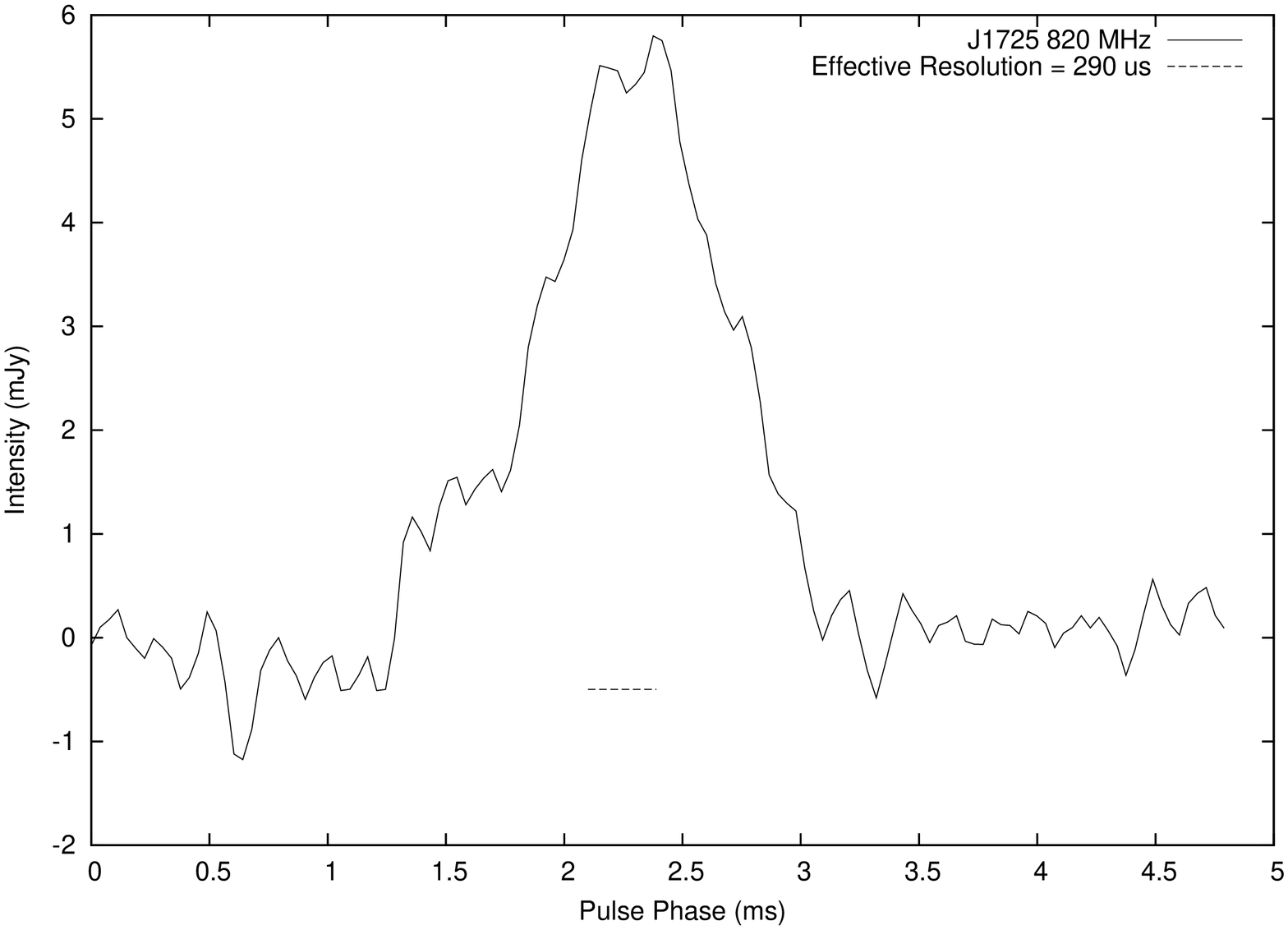} &
\includegraphics[scale=0.25, angle=-90]{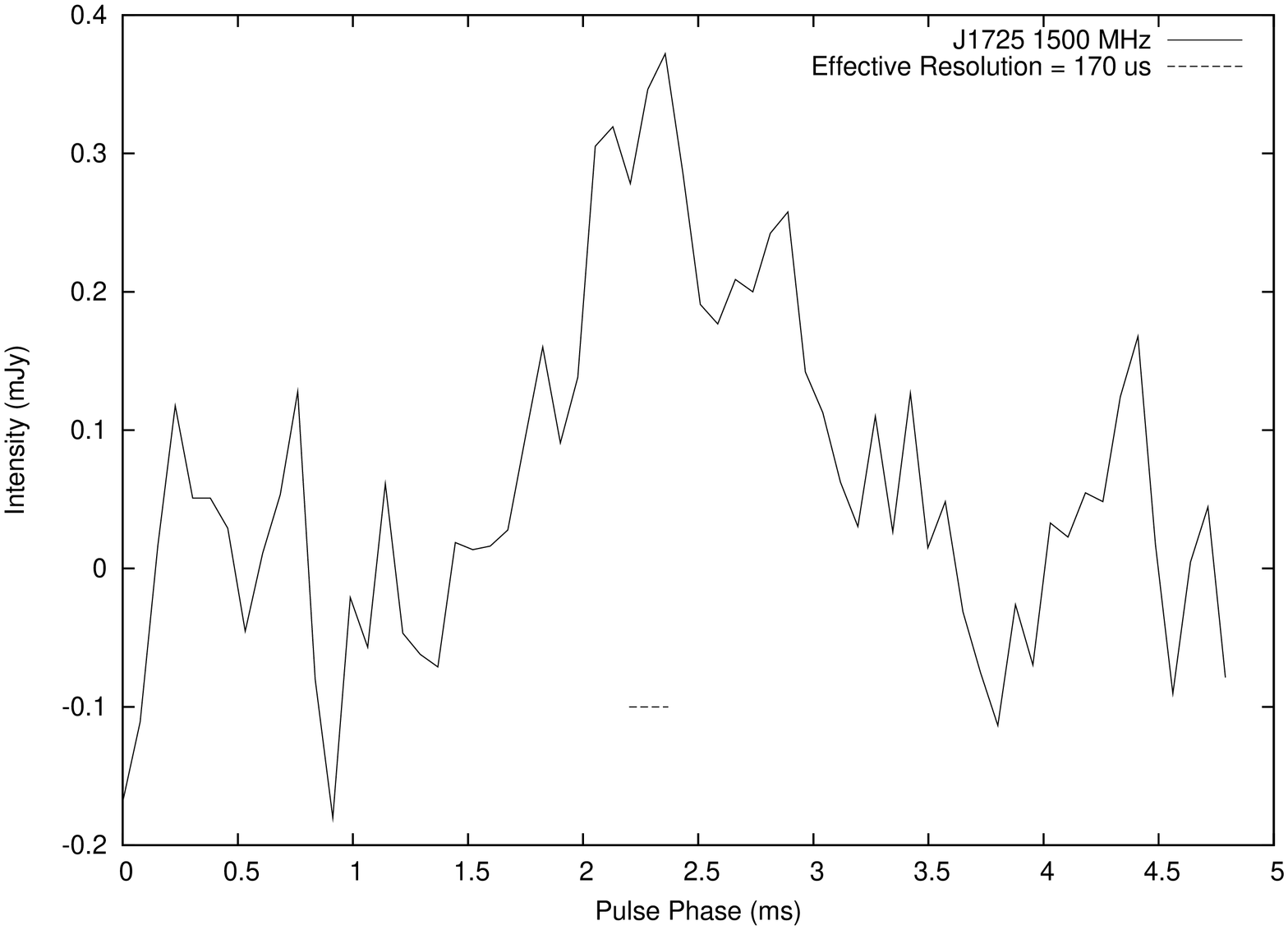} \\
\end{tabular}
\caption{{\it Upper Left:} Folded profile from the 35 minute confirmation observation of PSR J0922$-$52 on MJD 56102 at 
1400 MHz. The effective resolution of the profile (given by $t_{eff} = \sqrt{t_{samp}^2 + t_{scatt}^2 +t_{DM}^2}$, where 
$t_{samp}$ is the sampling time, $t_{scatt}$ is the scattering time from the NE2001 model \citep{Cordes:2002}, and $t_{DM}$ 
is the DM smearing across a single channel) is given in the plot and is shown by the bar beneath the profile. {\it Upper 
Right:} Folded profile from the 20 minute confirmation observation of PSR J1147$-$66 on MJD 56158 at 1400 MHz. {\it Middle 
Left:} Folded profile from the 15 minute confirmation observation of PSR J1227$-$6208 on MJD 55857 at 1400 MHz. {\it Middle 
Right:} Folded profile from the 35 minute confirmation observation of PSR J1546$-$59 on MJD 56102 at 1400 MHz. {\it Lower 
Left:} Composite profile for PSR J1725$-$3853 at 820 MHz, with a total integration time of 113 minutes. {\it Lower Right:} 
Folded profile from the 15 minute 1500 MHz observation of PSR J1725$-$3853 on MJD 55876.}
\label{fig:comp_profs}
\end{figure}

PSR J0922$-$52 has a period of 9.68~ms and a DM of 122.4~pc cm$^{-3}$. The spectral S/N and $\chi^2$ of the profile from 
the discovery observation, reported by {\tt seek} and {\tt prepfold}, are 9.1 and 2.7, respectively. The inferred distance 
from the NE2001 model \citep{Cordes:2002} is 0.8~kpc. It was confirmed on MJD 56102 with a 35 minute observation using the 
Parkes telescope at 1400 MHz. The full width at half maximum (FWHM) of the profile from the confirmation observation 
(Figure \ref{fig:comp_profs}) is 790~$\mu$s. Since the position is not well constrained by the observations we have been 
able to carry out, we can only estimate, using the radiometer equation \citep{Lorimer:2005}, a lower limit on the mean flux 
at 1400 MHz, which is 0.16~mJy. Further observations are needed in order to time this pulsar and determine its physical 
parameters.

\subsection{PSR J1147$-$66}
\label{1147}

PSR J1147$-$66 has a period of 3.72~ms and a DM of 133.8~pc cm$^{-3}$. The spectral S/N and $\chi^2$ of the profile from 
the discovery observation, reported by {\tt seek} and {\tt prepfold}, are 10.9 and 6.0, respectively. The inferred distance 
from the NE2001 model is 2.7~kpc. It was confirmed on MJD 56158 with a 20 minute observation using the Parkes telescope at 
1400 MHz. The FWHM of the profile from the confirmation observation (Figure \ref{fig:comp_profs}) is 795~$\mu$s. The 
estimate of the lower limit on the mean flux at 1400 MHz is 0.80~mJy. As with PSR J0922$-$52, further observations are 
needed for timing and determining its physical characteristics.

\subsection{PSR J1227$-$6208}
\label{1227}

PSR J1227$-$6208 has a period of 34.53~ms and a DM of 362.6~pc cm$^{-3}$. The spectral S/N and $\chi^2$ of the profile from 
the discovery observation, reported by {\tt seek} and {\tt prepfold}, are 12.6 and 4.6, respectively. The inferred distance 
from the NE2001 model is 8.3~kpc. It was confirmed on MJD 55857 with a 15 minute observation using the 
Parkes telescope at 1400 MHz, and was independently detected by the High Time Resolution Universe (HTRU) pulsar survey 
\citep{Keith:2010}, as well as the ongoing processing by Einstein@Home (Knispel et al., in prep). A full timing solution 
will be given by Thornton et al.\ (in prep), who found it to be in an approximately circular binary orbit of period 6.7 
days with a $\gtrsim$1.3 $M_{\odot}$ companion. Since a companion of this mass could be a neutron star, we searched both 
the original PMPS data and our confirmation observation for another pulsar, but found none down to a flux limit of 
0.16~mJy, assuming a detection significance of 6$\sigma$. Unlike previous searches of this kind 
\citep[e.g.][]{Lorimer:2006}, no correction for acceleration is needed, as the orbital period is substantially longer than 
the survey integration time (35 minutes). The FWHM of the profile from the confirmation observation (Figure 
\ref{fig:comp_profs}) is 1.3~ms. The estimate of the lower limit on the mean flux at 1400 MHz is 0.27~mJy. Further details 
of this pulsar will be published by Thornton et al.\ (in prep).

\subsection{PSR J1546$-$59}
\label{1546}

PSR J1546$-$59 has a period of 7.80~ms and a DM of 168.3~pc cm$^{-3}$. The spectral S/N and $\chi^2$ of the profile from 
the discovery observation, reported by {\tt seek} and {\tt prepfold}, are 9.4 and 3.4, respectively. The inferred distance 
from the NE2001 model is 3.3~kpc. It was confirmed on MJD 56102 with a 35 minute observation using the Parkes telescope at 
1400 MHz. The FWHM of the profile from the confirmation observation (Figure \ref{fig:comp_profs}) is 670~$\mu$s. The 
estimate of the lower limit on the mean flux at 1400 MHz is 0.20~mJy. As with PSRs J0922$-$52 and J1147$-$66, further 
observations are needed for timing and determining its physical characteristics.

\subsection{PSR J1725$-$3853}
\label{1725}

PSR J1725$-$3853 has a period of 4.79~ms and a DM of 158.2~pc cm$^{-3}$. The spectral S/N and $\chi^2$ of the profile from 
the discovery observation, reported by {\tt seek} and {\tt prepfold}, are 10.2 and 4.6, respectively. The inferred distance 
from the NE2001 model is 2.8~kpc. Both the confirmation observation on MJD 55660 and timing observations were done with the 
100-m Robert C.\ Byrd Green Bank Telescope (GBT) at 820 MHz, with one timing observation at 1500 MHz. The observational 
parameters are listed in Table \ref{table:obs}. All observations were taking using GUPPI, the Green Bank Ultimate Pulsar 
Processing Instrument \citep{DuPlain:2008}, which is built from reconfigurable off-the-shelf hardware and software available 
from CASPER \citep[Collaboration for Astronomy Signal Processing and Electronics Research;][]{Parsons:2009}. GUPPI samples 
data with 8-bit precision over bandwidths as large as 800 MHz, and is capable of recording all four Stokes parameters. The 
observations taken on MJDs 55825 and 56031 were done in a gridding format \citep[see, e.g.,][]{Morris:2002}, where we took 
four observations around the position of the pulsar in order to better constrain the position, which greatly facilitated 
the timing analysis described below.

The GBT data were initially optimized by a fine search in period and dispersion measure to produce integrated pulse 
profiles. Using a simple Gaussian template, we extracted times-of-arrival (TOAs) from each profile via the Fourier-domain 
template matching algorithm \citep{Taylor:1992} as implemented in the {\tt get\_toa.py} routine in the PRESTO package. The 
Gaussian template for the 820 MHz observations was made from the composite 820 MHz profile, and the template for the 1500 
MHz observation was made from the one observation at that frequency. Gaussian templates were used becuase a template made 
from the composite profile underestimated the errors on the residuals by a factor of nine. In total, a set of 13 TOAs 
spanning 371 days were then fit to a simple isolated pulsar timing model using the TEMPO analysis 
package\footnote{\url{http://tempo.sourceforge.net}}. Following a number of iterations, we were able to converge on a 
timing model in which the TOAs are fit by an isolated pulsar with parameters listed in Table \ref{table:timing}. The TOA 
uncertainties were multiplied by a factor of 3 to ensure a reduced $\chi^2$ value in the fit of unity. The root-mean-square 
timing model residuals were 88~$\mu$s. The positional uncertainty resulting from this fit is 0.05" in declination and 1.2" 
in RA, while the frequency derivative is only marginally significant, given the current baseline. The final fit parameters 
are typical for an isolated milliscond pulsar with a characteristic age of approximately 1.6~Gyr and a surface magnetic 
field of 5$\times$10$^8$~Gauss \citep[see, e.g.,][]{Lorimer:2008}.

No coincident sources were detected in any HEASARC catalogue\footnote{\url{http://heasarc.gsfc.nasa.gov}}, and no 
$\gamma$-ray source was detected in 371 days of folded $\gamma$-ray photons from the \emph{Fermi} Large Area Telescope 
\citep{Atwood:2009}. The FWHM of the composite profile made from adding all of the profiles at 820 MHz is 865~$\mu$s, and 
the FWHM of the 1500 MHz profile is 1.2~ms (Figure \ref{fig:comp_profs}). The estimated mean flux at 820 MHz is 1.1~mJy. We 
were unable to compute a reliable flux density at 1500 MHz given the 0.15~mJy detection threshold of the PMPS 
\citep{Manchester:2001}. These flux estimates have large errors, and further observations are required to reliably 
calculate a spectral index.

In addition to the five pulsars confirmed so far, the search analysis described in Section \ref{data} resulted in a 
large number of statistically significant candidate pulsar signals. A list of these candidates, which will be subject to 
follow-up observations with the GBT and Parkes, can be found at \url{http://astro.phys.wvu.edu/pmps}.

\section{Conclusions}
\label{conc}

Reprocessing of the Parkes Multibeam Pulsar Survey resulted in the discovery and confirmation of five new pulsars, PSR 
J0922$-$52, PSR J1147$-$66, PSR J1227$-$6208, PSR J1546$-$59, and PSR J1725$-$3853. PSR J1227$-$6208 was independently 
confirmed by Einstein@Home as well as the HTRU team in their medium-latitude survey and will be presented by Thornton et 
al.\ (in prep). Our discovery of PSRs J0922$-$52, J1147$-$66, J1227$-$6208, J1546$-$59, and J1725$-$3853 brings the total 
number of millisecond pulsars found in the PMPS to 25. We present a timing solution for PSR J1725$-$3853, and continued 
timing observations will allow us to further improve this solution.

Our discovery of these five pulsars emphasizes the value of archiving pulsar search data and indicates that there are a 
number of as-yet-undiscovered pulsars present in the PMPS data. Given the number of pulsar candidates present, automated 
searches are the most efficient way to reduce the number of candidates to an amount that can be viewed in a reasonable 
amount of time. Due to the fact that they have both high DMs and short periods, many of our candidates are weak and close 
to the detection threshold, so there is a good chance they were not ranked highly by previous automated searches. 
\citet{Keith:2009} found that weak pulsars and pulsars with high DMs were ranked highly by automated searches. However, 
most of these pulsars have long periods, i.e.\ periods on the order of hundreds of milliseconds. As the ratio of DM to 
period increases, the detected pulse profile is significantly broadened and begins to look more sinusoidal. These 
candidates are harder to select via ranking systems. We note that \citet{Eatough:2010} found that artificial neural 
networks have difficulty detecting short period pulsars, with their own detecting only 50\% of pulsars with periods less 
than 10~ms. In our search strategy, every single candidate is being inspected by eye. In many of the earlier analyses of 
the PMPS data (e.g.~Manchester et al.~2001), the candidates were also viewed by eye and it is not clear why these were not 
found earlier. Perhaps they were simply missed due to human fatigue. In the following year, we hope to follow up and 
confirm many of our candidates. Along with the re-analysis of the PMPS survey data presented here, and the ongoing search 
by Einstein@Home, we expect the sample of millisecond pulsars found in the PMPS to increase further.

\section*{Acknowledgements}

We would like to thank the HTRU team and in particular Michael Keith, Matthew Bailes, Sam Bates, and Dan Thornton for their 
information on the binary parameters of PSR J1227$-$6208. We would also like to thank Matthew Bailes for making {\tt 
dedisperse\_all} available to us. We would like to thank Bruce Allen and Michael Kramer for directing our attention to the 
Einstein@Home webpage. We particularly thank Ben Knispel for very useful discussions about the Einstein@Home discoveries. 
We thank George Hobbs and Fernando Camilo for useful discussions concerning timing of PSR J1725$-$3853, and the referee for 
very helpful comments on the manuscript. The National Radio Astronomy Observatory is a facility of the National Science 
Foundation operated under cooperative agreement by Associated Universities, Inc. Pulsar research at WVU is supported by a 
WVEPSCoR research challenge grant awarded to M.A.M and D.R.L. In addition, A.C., L.H., and C.T. were supported by a 
Cottrell Scholar award to DRL.

\clearpage

\clearpage

\begin{table}[ht]
\centering
\caption{Observational parameters for the confirmation and timing observations of PSR J1725$-$3853.}
\medskip
\begin{tabular}{cccc}
\hline
MJD & Frequency (MHz) & Bandwidth (MHz) & Sampling Time ($\mu$s) \\
\hline
55660 & 820 & 200 & 81.92 \\
55707 & 820 & 200 & 81.92 \\
55780 & 820 & 200 & 81.92 \\
55825 & 820 & 200 & 81.92 \\
55876 & 1500 & 800 & 81.92 \\
56005 & 820 & 200 & 40.96 \\
56031 & 820 & 200 & 40.96 \\
\hline
\end{tabular}
\label{table:obs}
\end{table}

\begin{table}[ht]
\centering
\caption{Timing and derived parameters for PSR J1725$-$3853. Errors quoted are twice the nominal values reported by TEMPO
and reflect the uncertainties in the least significant digit.}
\medskip
\begin{tabular}{lc}
\hline
\multicolumn{2}{c}{Timing Parameters} \\
\hline
Right Ascension (J2000) & 17:25:27.27(8) \\
Declination (J2000) & $-$38:53:04.20(5) \\
Spin Period (s) & 0.004791822704(3) \\
Period Derivative (s s$^{-1}$) & 5(3)$\times$10$^{-20}$ \\
Dispersion Measure (pc cm$^{-3}$) & 158.2(7) \\
Reference Epoch (MJD) & 55846 \\
Number of TOAs & 13 \\
Span of Timing Data & 55660$-$56031 \\
\hline
\multicolumn{2}{c}{Derived Parameters} \\
\hline
Galactic Longitude (degrees) & 349.3(7) \\
Galactic Latitude (degrees) & $-$1.8(6) \\
Distance (kpc) & 2.8 \\
Surface Magnetic Field (Gauss) & 5(1)$\times$10$^{8}$ \\
Spin Down Luminosity (ergs s$^{-1}$) & 6(3)$\times$10$^{33}$ \\
Characteristic Age (yr) & 1.6(9)$\times$10$^{9}$ \\
820 MHz Flux Density (mJy) & 1.1 \\
Pulse FWHM at 820 MHz (ms) & 0.865 \\
Pulse FWHM at 1500 MHz (ms) & 1.2 \\
\hline
\end{tabular}
\label{table:timing}
\end{table}


\begin{thebibliography}{}

\bibitem[Atwood et al.(2009)]{Atwood:2009} Atwood, W.~B., et al.\
2009, \apj, 697, 1071

\bibitem[Cordes \& Lazio(2002)]{Cordes:2002} Cordes, J.~M.,
\& Lazio, T.~J.~W.\ 2002, arXiv:astro-ph/0207156

\bibitem[DuPlain et al.(2008)]{DuPlain:2008} DuPlain, R.,
Ransom, S., Demorest, P., et al.\ 2008, \procspie, 7019, 70191D

\bibitem[Eatough et al.(2009)]{Eatough:2009} Eatough, R.~P., Keane,
E.~F., \& Lyne, A.~G.\ 2009, \mnras, 395, 410

\bibitem[Eatough et al.(2010)]{Eatough:2010} Eatough, R.~P.,
Molkenthin, N., Kramer, M., et al.\ 2010, \mnras, 407, 2443

\bibitem[Eatough et al.(2011)]{Eatough:2011} Eatough, R.~P., Kramer,
M., Lyne, A.~G., \& Keith, M.\ 2011, American Institute of Physics 
Conference Series, 1357, 58

\bibitem[Faulkner et al.(2004)]{Faulkner:2004} Faulkner, A.~J.,
Stairs, I.~H., Kramer, M., et al.\ 2004, \mnras, 355, 147

\bibitem[Hobbs et al.(2004)]{Hobbs:2004} Hobbs, G., Faulkner, A.,
Stairs, I.~H., et al.\ 2004, \mnras, 352, 1439

\bibitem[Keane et al.(2010)]{Keane:2010} Keane, E.~F., Ludovici,
D.~A., Eatough, R.~P., et al.\ 2010, \mnras, 401, 1057

\bibitem[Keane et al.(2011)]{Keane:2011} Keane, E.~F., Kramer, M.,
Lyne, A.~G., Stappers, B.~W., \& McLaughlin, M.~A.\ 2011, \mnras, 415, 3065

\bibitem[Keith et al.(2009)]{Keith:2009} Keith, M.~J., Eatough,
R.~P., Lyne, A.~G., et al.\ 2009, \mnras, 395, 837

\bibitem[Keith et al.(2010)]{Keith:2010} Keith, M.~J., Jameson,
A., van Straten, W., et al.\ 2010, \mnras, 409, 619

\bibitem[Kramer et al.(2003)]{Kramer:2003} Kramer, M., Bell, J.~F.,
Manchester, R.~N., et al.\ 2003, \mnras, 342, 1299

\bibitem[Lorimer \& Kramer(2005)]{Lorimer:2005} Lorimer, D. R. 
\& Kramer, M.\ 2005, Handbook of Pulsar Astronomy

\bibitem[Lorimer et al.(2006)]{Lorimer:2006} Lorimer, D.~R.,
Faulkner, A.~J., Lyne, A.~G., et al.\ 2006, \mnras, 372, 777

\bibitem[Lorimer(2008)]{Lorimer:2008} Lorimer, D.~R.\ 2008, Living
Reviews in Relativity, 11, 8

\bibitem[Manchester et al.(2001)]{Manchester:2001} Manchester, R.~N.,
Lyne, A.~G., Camilo, F., et al.\ 2001, \mnras, 328, 17

\bibitem[McLaughlin et al.(2006)]{McLaughlin:2006} McLaughlin, M.~A.,
Lyne, A.~G., Lorimer, D.~R., et al.\ 2006, \nat, 439, 817

\bibitem[Morris et al.(2002)]{Morris:2002} Morris, D.~J., Hobbs,
G., Lyne, A.~G., et al.\ 2002, \mnras, 335, 275

\bibitem[Parsons et al.(2009)]{Parsons:2009} Parsons, A., et al.\
2009, arXiv:0904.1181v1

\bibitem[Ransom et al.(2002)]{Ransom:2002} Ransom, S.~M.,
Eikenberry, S.~S., \& Middleditch, J.\ 2002, \aj, 124, 1788

\bibitem[Taylor(1992)]{Taylor:1992} Taylor, J.~H.\ 1992, Royal
Society of London Philisophical Transactions Series A, 341, 117

\end{thebibliography}
\end{document}